\documentclass{article}%
\usepackage{amssymb}
\usepackage{amsfonts}
\usepackage{amsmath}
\usepackage{graphicx}%
\providecommand{\U}[1]{\protect\rule{.1in}{.1in}}
\providecommand{\U}[1]{\protect\rule{.1in}{.1in}}

\begin{document}

\title{$SU(1,1)$ group action and the corresponding Bloch equation}
\author{A. Okninski\\Physics Division, Politechnika Swietokrzyska,\\Al. 1000-lecia PP 7, 25-314 Kielce, Poland.}
\maketitle

\begin{abstract}
Discrete time dynamics on the $SU\left(  1,1\right)  $ group is studied. It is
shown that a map acting in the dual space is the stroboscopic map for a
$SU\left(  1,1\right)  $ Bloch equation. Exact solution of the map is used to
elucidate the corresponding dynamics. It is shown that dynamics of the 
$SU\left(  1,1\right)  $\ Bloch equation
in the elliptic case bears close analogy to the $SU\left(  2\right)  $ Bloch dynamics.

\end{abstract}

\section{Introduction}

Discrete-time dynamical systems can be formulated in terms of group actions to
exploit the group structure and get a better understanding of the
corresponding dynamics. This approach was used to study discrete-time dynamics
on some groups, see \cite{Okninski2009} and references therein. On the other
hand, structure of Kleinian groups is naturally studied in the setting of
discrete-time dynamical systems, revealing in this way connections with
fractals \cite{Brooks1981,Gehring1989,Mumford2002}. For example, the
Shimizu-Leutbecher map is a typical tool to study group structure
\cite{Shimizu1963,Leutbecher1967}, see also
\cite{Brooks1981,Gehring1989,Beardon1983}.

Recently, we have investigated a possibility of relating group actions with
stroboscopic maps of ordinary differential equations \cite{Okninski2009}. More
exactly, we have studied the following dynamical system on a Lie group
$\mathcal{G}$:%
\begin{equation}
R_{N+1}=Q_{N}R_{N}Q_{N-1}R_{N-1}Q_{N-1}^{-1}R_{N}^{-1}Q_{N}^{-1},\qquad
N=1,2,\ldots, \label{Q_NR_N}%
\end{equation}
where $Q_{N},\ R_{N}\in\mathcal{G}$. A general solution of the map
(\ref{Q_NR_N}) has been constructed and it was demonstrated that\ for
$\mathcal{G}=SU\left(  2\right)  $ and $Q_{N}\equiv Q$\ Eq.(\ref{Q_NR_N}) is a
stroboscopic map of the Bloch equation \cite{Okninski2009}. The latter result
is generalized in the present paper for the case $\mathcal{G}=SU\left(
1,1\right)  $.

Let us note here that the $SU\left(  1,1\right)  $ Bloch equation finds
important applications in quantum optics. More exactly, the group of
squeezings is generated by the Lie algebra $\mathfrak{su}\left(  1,1\right)
$, the geometry of group manifold is that of Minkowski spaces and time
evolution is described by the $SU\left(  1,1\right)  $ Bloch equation
\cite{Dattoli1986, Aravind1988, King1999, Puri2001}.

The paper is organized as follows. In the next Section discrete time dynamics,
defined and solved in \cite{Okninski2009} for arbitrary group $\mathcal{G}$,
is studied in the case $\mathcal{G}=SU\left(  1,1\right)  $. It is shown in
Section 3 that the map (\ref{Q_NR_N}), considered in the dual space, is the
stroboscopic map for the $SU\left(  1,1\right)  $ Bloch equation which is
written in the elliptic, parabolic and hyperbolic cases. In Section 4
computations intended to elucidate dynamics of the $SU\left(  1,1\right)
$\ Bloch equation are presented for the elliptic case and analogy with the
$SU\left(  2\right)  $ Bloch equation is stressed. The obtained results are
summarized in the last Section.

\section{Discrete-time dynamics on the $SU\left(  1,1\right)  $ group}

Let us recall that the map (\ref{Q_NR_N}) admits an exact solution:
\begin{subequations}
\label{SOL}%
\begin{align}
R_{2K}  &  =S_{2K}S_{2K-1}\ldots S_{2}R_{0}S_{2}^{-1}\ldots S_{2K-1}%
^{-1}S_{2K}^{-1},\label{solR}\\
S_{N}  &  =Q_{N-1}\ldots Q_{1}S_{1}Q_{-1}^{-1}\ldots Q_{N-3}^{-1},
\label{solS}%
\end{align}
where
\end{subequations}
\begin{equation}
S_{N}\overset{df}{=}R_{N}Q_{N-1}R_{N-1}Q_{N-2}, \label{defS}%
\end{equation}
and similar equations can be written for $R_{2K+1}$ \cite{Okninski2009}.

We shall consider a special case $Q_{N}\equiv Q$ in (\ref{Q_NR_N}). In the
case $\mathcal{G}=SU(1,1)$ the following parameterization is used
\cite{Chiribella2006}:%
\begin{align}
R_{N}  &  =\exp\left(  i\tfrac{\chi_{N}}{2}\overrightarrow{\kappa}%
\cdot\overrightarrow{r_{N}}\right)  ,\qquad\overrightarrow{r_{N}}%
\cdot\overrightarrow{r_{N}}=\eta,\label{defR}\\
Q  &  =\exp\left(  i\tfrac{\alpha}{2}\overrightarrow{\kappa}\cdot
\overrightarrow{q}\right)  ,\hspace{0.4in}\overrightarrow{q}\cdot
\overrightarrow{q}=\eta, \label{defQ}%
\end{align}
where $i^{2}=-1$, $\eta\in\left\{  +1,0,-1\right\}  $ and is fixed,
$\overrightarrow{\kappa}\overset{df}{=}\left[  i\sigma^{1},i\sigma^{2}%
,\sigma^{3}\right]  $ where $\sigma^{1}$, $\sigma^{2}$, $\sigma^{3}$ are the
Pauli matrices\ and scalar products are defined as
\begin{subequations}
\label{DEFSCALAR}%
\begin{align}
&  \overrightarrow{\kappa}\cdot\overrightarrow{x}\overset{df}{=}-\kappa
^{1}x^{1}-\kappa^{2}x^{2}+\kappa^{3}x^{3},\label{defkx}\\
&  \overrightarrow{x}\cdot\overrightarrow{x}\overset{df}{=}-x^{1}x^{1}%
-x^{2}x^{2}+x^{3}x^{3}, \label{defxx}%
\end{align}
where $\overrightarrow{x}\overset{df}{=}\left[  x^{1},x^{2},x^{3}\right]  $.
The three cases $\eta=+1,0,-1$ are referred to as elliptic, parabolic and
hyperbolic, respectively. For an exposition of theory of the $SU(1,1)$ group
the reader can consult \cite{Chiribella2006, Wawrzynczyk1984}.

We obtain from (\ref{SOL}) the following solution:
\end{subequations}
\begin{equation}
R_{2K}=Q^{2K}P^{K}R_{0}P^{-K}Q^{-2K},\label{2k}%
\end{equation}
where $P\overset{df}{=}Q^{-1}S_{1}Q^{-1}=Q^{-1}R_{1}QR_{0}
,\ K=0,\ 1,\ 2,\ \ldots\ $. We still have to impose initial condition $R_{0}$
while $R_{1}$ is computed as $R_{1}=QPR_{0}^{-1}Q^{-1}$.

Matrix $P$ is parameterized in form%
\begin{equation}
P=\exp\left(  i\tfrac{\beta}{2}\overrightarrow{\kappa}\cdot\overrightarrow
{p}\right)  ,\qquad\overrightarrow{p}\cdot\overrightarrow{p}=\eta, \label{P}%
\end{equation}
and equation (\ref{2k}) can be written as%
\begin{equation}
\overrightarrow{\kappa}\cdot\overrightarrow{r}_{2K}=\exp\left(  iK\alpha
\overrightarrow{\kappa}\cdot\overrightarrow{q}\right)  \exp\left(
iK\tfrac{\beta}{2}\overrightarrow{\kappa}\cdot\overrightarrow{p}\right)
\,\overrightarrow{\kappa}\cdot\overrightarrow{r}_{0}\,\exp\left(
-iK\tfrac{\beta}{2}\overrightarrow{\kappa}\cdot\overrightarrow{p}\right)
\exp\left(  -iK\alpha\overrightarrow{\kappa}\cdot\overrightarrow{q}\right)
\mathbf{,} \label{RSsol1a}%
\end{equation}
and, after introducing new quantities, $K\alpha=\theta,\ \lambda=\frac{\beta
}{2\alpha},$ reads
\begin{equation}
\overrightarrow{\kappa}\cdot\overrightarrow{r}\left(  \theta\right)
=\exp\left(  i\theta\overrightarrow{\kappa}\cdot\overrightarrow{q}\right)
\exp\left(  i\lambda\theta\overrightarrow{\kappa}\cdot\overrightarrow
{p}\right)  \,\overrightarrow{\kappa}\cdot\overrightarrow{r}_{0}\,\exp\left(
-i\lambda\theta\overrightarrow{\kappa}\cdot\overrightarrow{p}\right)
\exp\left(  -i\theta\overrightarrow{\kappa}\cdot\overrightarrow{q}\right)
\mathbf{,} \label{RSsol1b}%
\end{equation}
where $\overrightarrow{r}\left(  \theta\right)  \overset{df}{=}\overrightarrow
{r}_{2K}$.

\subsection{Elliptic case}

To obtain $\overrightarrow{r}\left(  \theta\right)  $ from Eq.(\ref{RSsol1b})
we shall need to compute $\overrightarrow{t}\left(  \gamma\right)  $ given by:%
\begin{equation}
\overrightarrow{\kappa}\cdot\overrightarrow{t}\left(  \gamma\right)
=S\ \overrightarrow{\kappa}\cdot\overrightarrow{t}\ S^{-1}=\exp\left(
i\tfrac{\gamma}{2}\overrightarrow{\kappa}\cdot\overrightarrow{s}\right)
\,\overrightarrow{\kappa}\cdot\overrightarrow{t}\,\exp\left(  -i\tfrac{\gamma
}{2}\overrightarrow{\kappa}\cdot\overrightarrow{s}\right)  , \label{deft}%
\end{equation}
in the case $\eta=1$. The exponential map is simplified as%
\begin{equation}
\exp\left(  i\tfrac{\gamma}{2}\overrightarrow{\kappa}\cdot\overrightarrow
{s}\right)  =\cos\left(  \tfrac{\gamma}{2}\right)  \mathbf{1}+i\sin\left(
\tfrac{\gamma}{2}\right)  \overrightarrow{\kappa}\cdot\overrightarrow
{s},\qquad\overrightarrow{s}\cdot\overrightarrow{s}=1. \label{expe}%
\end{equation}

Now, due to properties of the Pauli matrices we obtain from (\ref{deft}):%
\begin{equation}
\overrightarrow{t}\left(  \gamma\right)  =\cos\left(  \gamma\right)
\overrightarrow{t}-\sin\left(  \gamma\right)  \overrightarrow{t}%
\times\overrightarrow{s}+\left(  1-\cos\left(  \gamma\right)  \right)  \left(
\overrightarrow{t}\cdot\overrightarrow{s}\right)  \overrightarrow{s},
\label{te}%
\end{equation}
where%
\begin{equation}
\overrightarrow{x}\times\overrightarrow{y}\overset{df}{=}\left[  -x_{2}%
y_{3}+x_{3}y_{2},-x_{3}y_{1}+x_{1}y_{3},x_{1}y_{2}-x_{2}y_{1}\right]  .
\label{defxy}%
\end{equation}

Using twice the equation (\ref{te}) in (\ref{RSsol1b}) we get:
\begin{subequations}
\label{EXPLE}%
\begin{align}
\overrightarrow{r}\left(  \theta\right)   &  =\cos\left(  2\theta\right)
\overrightarrow{t}\left(  \theta\right)  -\sin\left(  2\theta\right)
\overrightarrow{t}\left(  \theta\right)  \times\overrightarrow{q}+\left(
1-\cos\left(  2\theta\right)  \right)  \left(  \overrightarrow{q}%
\cdot\overrightarrow{t}\left(  \theta\right)  \right)  \overrightarrow
{q}\mathbf{,}\label{r1e}\\
\overrightarrow{t}\left(  \theta\right)   &  =\cos\left(  2\lambda
\theta\right)  \overrightarrow{r}_{0}-\sin\left(  2\lambda\theta\right)
\overrightarrow{r}_{0}\times\overrightarrow{p}+\left(  1-\cos\left(
2\lambda\theta\right)  \right)  \left(  \overrightarrow{p}\cdot\overrightarrow
{r}_{0}\right)  \overrightarrow{p}. \label{r2e}%
\end{align}
For growing $\theta$ the vector $\overrightarrow{r}\left(  \theta\right)  $
evolves on manifold $\overrightarrow{r}\left(  \theta\right)  \cdot
\overrightarrow{r}\left(  \theta\right)  =1$, i.e. on two-sheeted hyperboloid.

\subsection{Parabolic case}

In the case $\eta=0$ the exponential map reduces to%
\end{subequations}
\begin{equation}
\exp\left(  i\tfrac{\gamma}{2}\overrightarrow{\kappa}\cdot\overrightarrow
{s}\right)  =\mathbf{1}+i\tfrac{\gamma}{2}\overrightarrow{\kappa}%
\cdot\overrightarrow{s},\qquad\overrightarrow{s}\cdot\overrightarrow{s}=0.
\label{expp}%
\end{equation}

Now, due to properties of the Pauli matrices we compute from (\ref{deft}):%
\begin{equation}
\overrightarrow{t}\left(  \gamma\right)  =\overrightarrow{t}-\gamma
\overrightarrow{t}\times\overrightarrow{s}+\left(  \overrightarrow{t}%
\cdot\overrightarrow{s}\right)  \overrightarrow{s}. \label{tp}%
\end{equation}
Using twice the equation (\ref{tp}) in (\ref{RSsol1b}) we obtain:%

\begin{subequations}
\label{EXPLP}%
\begin{align}
\overrightarrow{r}\left(  \theta\right)   &  =\overrightarrow{t}\left(
\theta\right)  -2\theta\overrightarrow{t}\left(  \theta\right)  \times
\overrightarrow{q}+\left(  \overrightarrow{q}\cdot\overrightarrow{t}\left(
\theta\right)  \right)  \overrightarrow{q}\mathbf{,}\label{r1p}\\
\overrightarrow{t}\left(  \theta\right)   &  =\overrightarrow{r}_{0}%
-2\lambda\theta\overrightarrow{r}_{0}\times\overrightarrow{p}+\left(
\overrightarrow{p}\cdot\overrightarrow{r}_{0}\right)  \overrightarrow{p}.
\label{r2p}%
\end{align}
The vector $\overrightarrow{r}\left(  \theta\right)  $ evolves on manifold
$\overrightarrow{r}\left(  \theta\right)  \cdot\overrightarrow{r}\left(
\theta\right)  =0$, i.e. on the Minkowski cone.

\subsection{Hyperbolic case}

In the case $\eta=-1$ we have%
\end{subequations}
\begin{equation}
\exp\left(  i\tfrac{\gamma}{2}\overrightarrow{\kappa}\cdot\overrightarrow
{s}\right)  =\cosh\left(  \tfrac{\gamma}{2}\right)  \mathbf{1}+i\sinh\left(
\tfrac{\gamma}{2}\right)  \overrightarrow{\kappa}\cdot\overrightarrow
{s},\qquad\overrightarrow{s}\cdot\overrightarrow{s}=-1. \label{exph}%
\end{equation}

Due to properties of the Pauli matrices we obtain from (\ref{deft}):%
\begin{equation}
\overrightarrow{t}\left(  \gamma\right)  =\cosh\left(  \gamma\right)
\overrightarrow{t}-\sinh\left(  \gamma\right)  \overrightarrow{t}%
\times\overrightarrow{s}+\left(  \cosh\left(  \beta\right)  -1\right)  \left(
\overrightarrow{t}\cdot\overrightarrow{s}\right)  \overrightarrow{s}.
\label{th}%
\end{equation}

Using twice the equation (\ref{th}) in (\ref{RSsol1b}) we get:
\begin{subequations}
\label{EXPLH}%
\begin{align}
\overrightarrow{r}\left(  \theta\right)   &  =\cosh\left(  2\theta\right)
\overrightarrow{t}\left(  \theta\right)  -\sinh\left(  2\theta\right)
\overrightarrow{t}\left(  \theta\right)  \times\overrightarrow{q}+\left(
\cosh\left(  2\theta\right)  -1\right)  \left(  \overrightarrow{q}%
\cdot\overrightarrow{t}\left(  \theta\right)  \right)  \overrightarrow
{q}\mathbf{,}\label{r1h}\\
\overrightarrow{t}\left(  \theta\right)   &  =\cosh\left(  2\lambda
\theta\right)  \overrightarrow{r}_{0}-\sinh\left(  2\lambda\theta\right)
\overrightarrow{r}_{0}\times\overrightarrow{p}+\left(  \cosh\left(
2\lambda\theta\right)  -1\right)  \left(  \overrightarrow{p}\cdot
\overrightarrow{r}_{0}\right)  \overrightarrow{p}. \label{r2h}%
\end{align}
The vector $\overrightarrow{r}\left(  \theta\right)  $ evolves on manifold
$\overrightarrow{r}\left(  \theta\right)  \cdot\overrightarrow{r}\left(
\theta\right)  =-1$, i.e. on one-sheeted hyperboloid.

\subsection{Symmetry and restrictions of dynamics}

Dynamical system (\ref{Q_NR_N}) for $Q_{N}\equiv Q$\ has continuous symmetry:%
\end{subequations}
\begin{equation}
R_{N}\rightarrow Q^{\kappa}R_{N}Q^{-\kappa},\qquad\forall\kappa\in\mathbb{R}.
\label{sym}%
\end{equation}
It can be thus expected that dynamics of the quantity $\overrightarrow
{r}\left(  \theta\right)  \cdot\overrightarrow{q}$ should decouple from other
degrees of freedom in (\ref{Q_NR_N}) \cite{Okninski2009}. Indeed, it follows
from (\ref{EXPLE}) that
\begin{equation}
\overrightarrow{r}\left(  \theta\right)  \cdot\overrightarrow{q}%
=\overrightarrow{t}\left(  \theta\right)  \cdot\overrightarrow{q}.
\label{decoupling}%
\end{equation}
Since in the elliptic or hyperbolic case $\overrightarrow{t}\cdot
\overrightarrow{t}=\overrightarrow{q}\cdot\overrightarrow{q}=\pm1$ it follows
from the Schwartz inequality for the Minkowski metric that $\left(
\overrightarrow{t}\cdot\overrightarrow{q}\right)  ^{2}\geq\left(
\overrightarrow{t}\cdot\overrightarrow{t}\right)  \left(  \overrightarrow
{q}\cdot\overrightarrow{q}\right)  =1$. Now, for given $\overrightarrow{p}$,
$\overrightarrow{q}$\ and $\overrightarrow{r}_{0}$\ on the upper sheet of the
hyperboloid in the elliptic case we have%
\begin{equation}
1\leq A_{1}\leq\overrightarrow{t}\left(  \theta\right)  \cdot\overrightarrow
{q}\leq A_{2}, \label{bounds1}%
\end{equation}
The constants $A_{1,2}$\ depending on the parameters $\overrightarrow{p}$,
$\overrightarrow{q}$\ and the initial condition $\overrightarrow{r}_{0}$\ can
be computed from (\ref{r2e}) by elementary means%
\begin{equation}
A_{1,2}=c\mp\sqrt{b^{2}+\left(  a-c\right)  ^{2}}, \label{bounds2}%
\end{equation}
where%
\begin{equation}
a=\overrightarrow{r}_{0}\cdot\overrightarrow{q}\mathbf{,\quad}b=\left(
\overrightarrow{r}_{0}\times\overrightarrow{p}\right)  \cdot\overrightarrow
{q}\mathbf{,\quad}c=\left(  \overrightarrow{p}\cdot\overrightarrow{r}%
_{0}\right)  \left(  \overrightarrow{p}\cdot\overrightarrow{q}\right)  .
\label{bounds3}%
\end{equation}
It thus follows that the motion on the hyperboloid\ is bounded by two
planes:\textsl{\ }$A_{1}\leq\overrightarrow{r}\left(  \theta\right)
\cdot\overrightarrow{q}\leq A_{2}$. In the parabolic case we obtain the
following simple condition%
\begin{equation}
\overrightarrow{t}\cdot\overrightarrow{q}=a-2\lambda\theta b+c, \label{cond}%
\end{equation}
i.e. equation of a straight line.

\section{The Bloch equation}

The map (\ref{Q_NR_N}), $Q_{N}\equiv Q$, is the stroboscopic map of a
differential equation which is conveniently deduced from the form
(\ref{RSsol1b}). Since $\alpha$ and $\beta$ are arbitrary we shall treat
$\theta$ as a continuous variable. Differentiating Eq.(\ref{RSsol1b}) with
respect to $\theta$ and using (\ref{RSsol1b}) we get
\begin{equation}
\dfrac{d\,\overrightarrow{\kappa}\cdot\overrightarrow{r}\left(  \theta\right)
}{d\theta}=i\left[  \overrightarrow{\kappa}\cdot\overrightarrow{u}\left(
\theta\right)  ,\ \overrightarrow{\kappa}\cdot\overrightarrow{r}\left(
\theta\right)  \right]  , \label{RSODE}%
\end{equation}
where $\left[  A,\ B\right]  \overset{df}{=}AB-BA$ and
\begin{subequations}
\label{DEFS}%
\begin{align}
\overrightarrow{u}\left(  \theta\right)   &  =\overrightarrow{q}%
+\lambda\overrightarrow{p}\left(  \theta\right)  ,\quad\left(  \lambda
=\beta/2\alpha\right) \label{defu}\\
\overrightarrow{\kappa}\cdot\overrightarrow{p}\left(  \theta\right)   &
=\exp\left(  i\theta\overrightarrow{\kappa}\cdot\overrightarrow{q}\right)
\,\overrightarrow{\kappa}\cdot\overrightarrow{p}\,\exp\left(  -i\theta
\overrightarrow{\kappa}\cdot\overrightarrow{q}\right)  . \label{defp}%
\end{align}

It follows that the sequence $\overrightarrow{r}_{0},\overrightarrow{r}%
_{2},\overrightarrow{r}_{4},\ldots,$ generated by $R_{0},R_{2},R_{4},\ldots,$
cf. Eq.(\ref{2k}), interpolates flow of Eq.(\ref{RSODE}). It turns out that, 
$\theta$ interpreted as time, Eq.(\ref{RSODE}) is the $SU\left(  1,1\right)  
$\ Bloch equation, cf.\cite{King1999}.

Equations (\ref{RSODE}), (\ref{DEFS}) can be written in explicit form.

\subsection{Elliptic case}

Using Eqs. (\ref{deft}), (\ref{te}) we get from (\ref{defp})
\end{subequations}
\begin{equation}
\overrightarrow{p}\left(  \theta\right)  =\cos\left(  \gamma\right)
\overrightarrow{p}-\sin\left(  \gamma\right)  \overrightarrow{p}%
\times\overrightarrow{q}+\left(  1-\cos\left(  \gamma\right)  \right)  \left(
\overrightarrow{p}\cdot\overrightarrow{q}\right)  \overrightarrow{q},
\label{defpe}%
\end{equation}

and using properties of the Pauli matrices we obtain from (\ref{RSODE}) the
Bloch equation in the elliptic case:%
\begin{equation}
\frac{d\,\overrightarrow{r}}{d\theta}=-2\overrightarrow{r}\left(
\theta\right)  \times\overrightarrow{u}\left(  \theta\right)  , \label{Bloche}%
\end{equation}
with $\overrightarrow{u}\left(  \theta\right)  $ and $\overrightarrow
{p}\left(  \theta\right)  $ given by (\ref{defu}) and (\ref{defpe}), respectively.

\subsection{Parabolic case}

Applying Eqs. (\ref{deft}), (\ref{tp}) to (\ref{defp})
\begin{equation}
\overrightarrow{p}\left(  \theta\right)  =\overrightarrow{p}-\gamma
\overrightarrow{p}\times\overrightarrow{q}+\left(  \overrightarrow{p}%
\cdot\overrightarrow{q}\right)  \overrightarrow{q}, \label{defpp}%
\end{equation}

and using properties of the Pauli matrices we obtain from (\ref{RSODE}) the
Bloch equation in the parabolic case:%
\begin{equation}
\frac{d\,\overrightarrow{r}}{d\theta}=-2\overrightarrow{r}\left(
\theta\right)  \times\overrightarrow{u}\left(  \theta\right)  , \label{Blochp}%
\end{equation}
with $\overrightarrow{u}\left(  \theta\right)  $ and $\overrightarrow
{p}\left(  \theta\right)  $ given by (\ref{defu}) and (\ref{defpp}), respectively.

\subsection{Hyperbolic case}

Using Eqs. (\ref{deft}), (\ref{th}) we obtain from (\ref{defp})
\begin{equation}
\overrightarrow{p}\left(  \theta\right)  =\cosh\left(  \gamma\right)
\overrightarrow{p}-\sinh\left(  \gamma\right)  \overrightarrow{p}%
\times\overrightarrow{q}+\left(  \cosh\left(  \gamma\right)  -1\right)
\left(  \overrightarrow{p}\cdot\overrightarrow{q}\right)  \overrightarrow{q},
\label{defph}%
\end{equation}

and using properties of the Pauli matrices we obtain from (\ref{RSODE}) the
Bloch equation in the hyperbolic case:%
\begin{equation}
\frac{d\,\overrightarrow{r}}{d\theta}=-2\overrightarrow{r}\left(
\theta\right)  \times\overrightarrow{u}\left(  \theta\right)  , \label{Blochh}%
\end{equation}
with $\overrightarrow{u}\left(  \theta\right)  $ and $\overrightarrow
{p}\left(  \theta\right)  $ given by (\ref{defu}) and (\ref{defph}), respectively.

\section{Computational results}

We have performed several computations for the Bloch equation (\ref{Bloche})
and the discrete-time dynamical system (\ref{Q_NR_N}), parameterized as in
Eqs. (\ref{defR}), (\ref{defQ}), $Q_{N}\equiv Q$, $\eta=1$. Exact solutions of
the map (\ref{Q_NR_N}) as well as of the Bloch equation (\ref{Bloche}) in the
elliptic case are given by (\ref{2k}) and (\ref{EXPLE}), respectively.

The solution (\ref{EXPLE}), of discrete-time dynamical system (\ref{Q_NR_N})
with $Q$, $P$ given by (\ref{defQ}), (\ref{P}) has been plotted in Fig.\ 1 for
$\overrightarrow{q}=\left[  0,0,1\right]  $, $\overrightarrow{p}=\left[
1,0,\sqrt{2}\right]  $, $\lambda=\beta/\left(  2\alpha\right)  =3$ and the
initial vector $\overrightarrow{r}_{0}=\left[  \frac{1}{2},\frac{1}{2}%
,\sqrt{\frac{3}{2}}\right]  $ on the upper sheet of the hyperboloid. The whole
trajectory has three-fold symmetry with respect to the $\overrightarrow{q}$
axis. Circles indicate parallels $A_{1,2}$ given by (\ref{bounds2}) confining
the dynamics.

In Fig. 2 dynamics of vectors $\overrightarrow{r}_{N}$ obtained from
(\ref{RSsol1a})\ has been plotted for $\alpha=5$, $\beta=20$ $\left(
\lambda=\beta/\left(  2\alpha\right)  =2\right)  $ and other parameters
unchanged. We thus obtain thirty six points marked with dots. The solution
(\ref{EXPLE}) has been also plotted. The closed curve has two-fold symmetry
with respect to the $\overrightarrow{q}$ axis.%

\begin{center}
\includegraphics[
trim=0.000000in 0.000000in -0.002699in 0.027395in,
height=3.0066in,
width=4.5338in
]%
{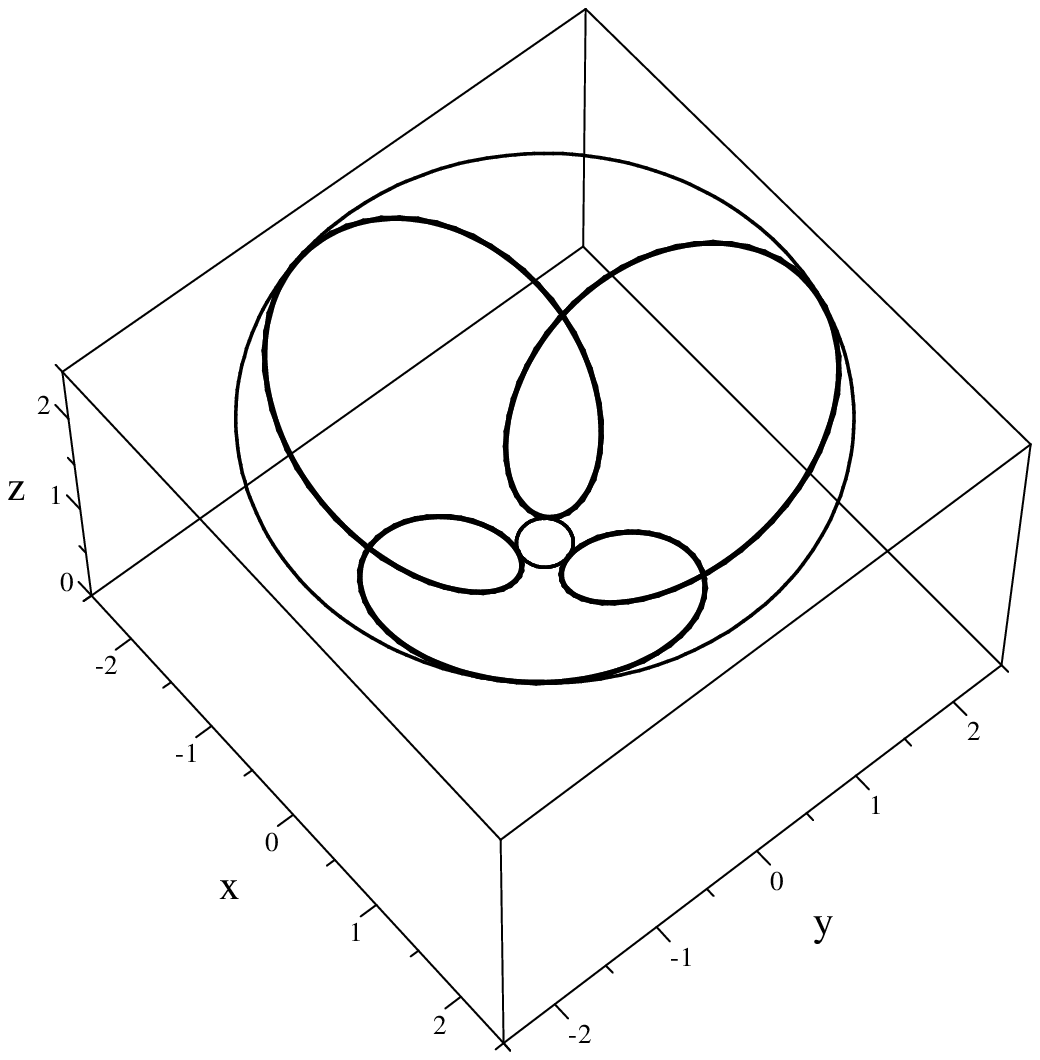}%
\\
Fig. 1. Exact solution of the Bloch equation (\ref{Bloche}), $\lambda=3$.
\end{center}
%

\begin{center}
\includegraphics[
height=3.0252in,
width=4.532in
]%
{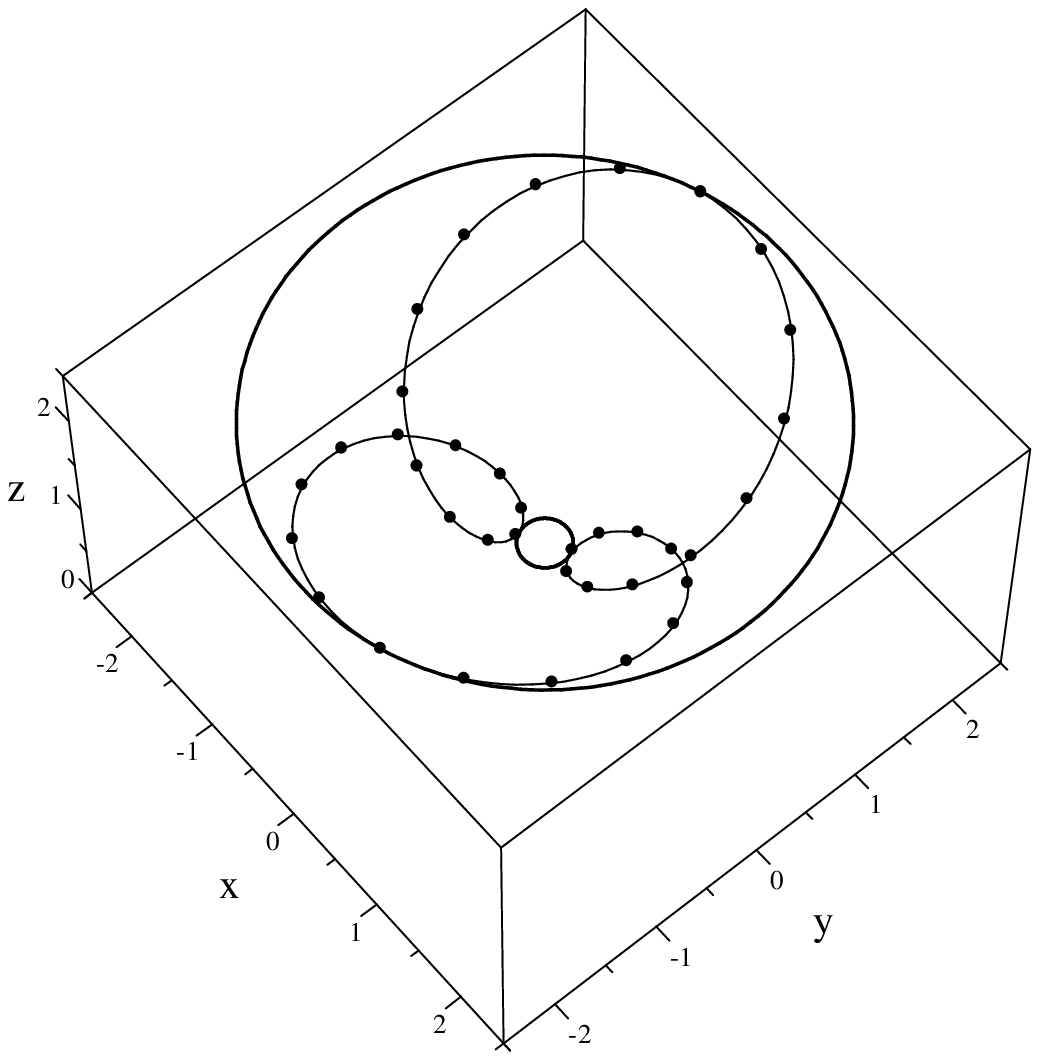}%
\\
Fig. 2. Exact solution of the Bloch equation (\ref{Bloche}) (thin line) and
discrete-time dynamical system (\ref{Q_NR_N}) (dots), $\alpha=5$, $\beta=20$,
$\lambda=2$.
\end{center}

In Fig. 3 initial stage of dynamics of vectors $\overrightarrow{r}_{N}$ has
been plotted for $\lambda=\beta/\left(  2\alpha\right)  =1.025$, dot marking
the initial vector $\overrightarrow{r}_{N}$.%

\begin{center}
\includegraphics[
height=3.0252in,
width=4.532in
]%
{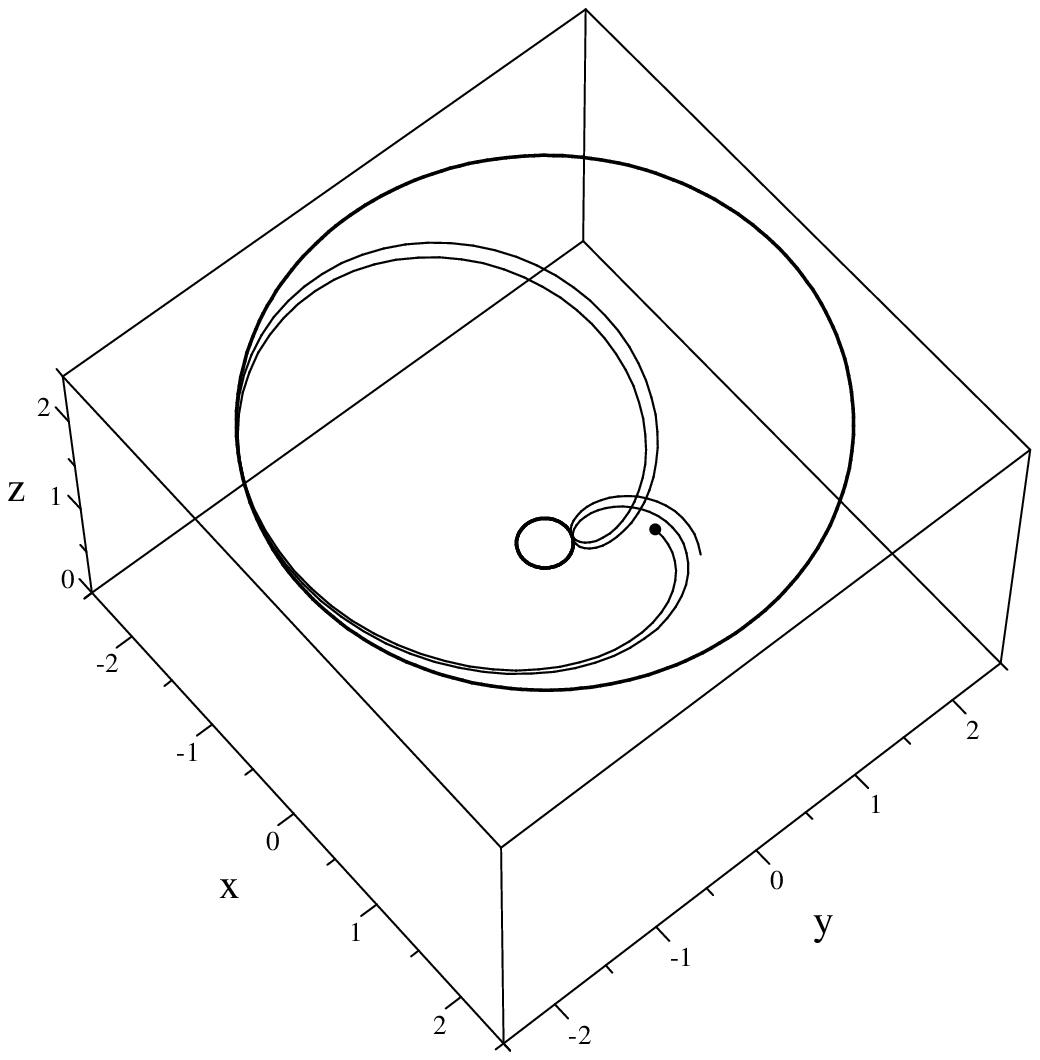}%
\\
Fig. 3. Exact solution of the Bloch equation (\ref{Bloche}), $\lambda=1.025$.
\end{center}

\section{Summary and discussion}

We have introduced in \cite{Okninski2009} a class of discrete-time invertible
maps (\ref{Q_NR_N}) on an arbitrary group $\mathcal{G}$ for which the exact
solution (\ref{SOL}) of this map has been found. Maps of form (\ref{Q_NR_N}),
parameterized on a Lie group, generate points in the dual (parameter) space
which sample a trajectory in this space arbitrarily densely. This curve can be
generated forward as well as backward from a given initial condition. This
suggests that the group action (\ref{Q_NR_N}) may correspond to a flow of a
differential equation. We have demonstrated in the present paper that for
$\mathcal{G} =SU\left(  1,1\right)  $ the map (\ref{Q_NR_N}), $Q_{N}\equiv Q$,
considered as dynamical system in the dual space, is a stroboscopic map of a
$SU\left(  1,1\right)  $ Bloch equation. It should be noted that the
$SU\left(  1,1\right)  $ Bloch equation (\ref{RSODE}) is formally analogous to
the $SU\left(  2\right)  $ Bloch equation, cf. Eq. (4.21) in
\cite{Okninski2009}.

Exact solutions of the map constructed in the present paper, (\ref{EXPLE}),
(\ref{EXPLP}), (\ref{EXPLH}), lead to a better understanding of the
corresponding Bloch equations (\ref{Bloche}), (\ref{Blochp}), (\ref{Blochh}).
More exactly, symmetries and restrictions of dynamics have been found
explicitely. It is interesting that dynamics of the $SU\left(  1,1\right)  $
Bloch equation in the elliptic case bears close analogy to dynamics of the
$SU\left(  2\right)  $ Bloch equation, compare Figs. 1, 2, 3 from the present
paper with analogous figures in \cite{Okninski2009}.

\end{document}